  \documentclass[a4paper, 12pt, pre, nofootinbib, onecolumn, groupedaddress, floatfix]{revtex4}
  \usepackage{amsmath, amstext, amssymb, amsfonts,amsmath}
  \usepackage{natbib,url,textcase,bm,color}
  \usepackage[dvips]{graphicx}
  \usepackage{epsfig}
  \usepackage[mathscr]{eucal}
  \usepackage{exscale}
  \usepackage[dvips, breaklinks, colorlinks=true,urlcolor=blue,citecolor=blue,linkcolor=blue,hyperindex]{hyperref}
  \usepackage[a4paper, top=2.3cm, bottom=1.1cm, right=1.5cm, left=1.5cm]{geometry}
  \usepackage{breakurl}
\graphicspath{ {./FIG/} }
\newcommand{\rd}{\mathrm{d}}
\newcommand{\vS}{\mathbf{S}}
\newcommand{\vL}{\mathbf{L}}
\newcommand{\de}[1]{\left(#1\right)}
\newcommand{\comu}[1]{\left[#1\right]}
\newcommand{\Ntil}{\widetilde{N}}

\newcommand{\media}[1]{\left\langle #1 \right\rangle}
\let \l = \left
\let \r = \right
\renewcommand{\baselinestretch}{1.15}
\begin{document}
%-----------------------------------------------------------------------
\title{Controlling the Range of Interactions in the Classical Inertial\vspace{0.13cm} Ferromagnetic Heisenberg Model: Analysis of Metastable States}
\renewcommand{\baselinestretch}{1.00}
%-----------------------------------------------------------------------
\author{Leonardo J.\ L.\ Cirto$^{1}$} \email{cirto@cbpf.br}
\author{Leonardo S.\ Lima$^{2}$}      \email{lslima@des.cefetmg.br}
\author{Fernando D.\ Nobre$^{1,3}$}   \email{fdnobre@cbpf.br}
%-----------------------------------------------------------------------
\affiliation{$^{1}$\footnotesize{Centro Brasileiro de Pesquisas F\'{\i}sicas, Rua Dr.\ Xavier Sigaud 150, 22290-180 Rio de Janeiro - RJ, Brazil}}
\affiliation{$^{2}$\footnotesize{Departamento de F\'\i sica e Matem\'atica, Centro Federal de Educa\c{c}\~ao Tecnol\'ogica de Minas Gerais, 30510-000 Belo Horizonte - MG, Brazil}}
\affiliation{$^{3}$\footnotesize{National Institute of Science and Technology for Complex Systems, Rua Dr.\ Xavier Sigaud 150, 22290-180 Rio de Janeiro - RJ,  Brazil}}
%-----------------------------------------------------------------------
\begin{abstract}
A numerical analysis of a one-dimensional Hamiltonian system, composed 
by~$N$ classical localized Heisenberg rotators on a ring, 
is presented. A distance~$r_{ij}$ between rotators at sites~$i$ and~$j$ is
introduced, such that the corresponding two-body interaction decays
with~$r_{ij}$ as a power-law, 
$1/r_{ij}^{\alpha}$ ($\alpha \ge 0$). 
The index~$\alpha$ controls the range of the interactions, in such a way that one recovers both the fully-coupled (i.e., mean-field limit)
and nearest-neighbour-interaction models in the particular limits $\alpha=0$ and $\alpha\to\infty$, respectively. 
The dynamics of the model is investigated for energies~$U$ below its critical value ($U<U_{c}$), with initial conditions corresponding to zero
magnetization. The presence of quasi-stationary states (QSSs), whose durations $t_{\rm QSS}$ increase for increasing values of~$N$, is verified for values
of~$\alpha$ in the range $0 \leq \alpha <1$, like the ones found for the similar model of XY rotators. Moreover, for a given energy~$U$,  our numerical analysis indicates 
that $t_{\rm QSS} \sim N^{\gamma}$, where the exponent $\gamma$ decreases for increasing~$\alpha$
in the range $0 \leq \alpha <1$, and particularly, our results suggest that $\gamma \to 0$ as $\alpha \to 1$. 
The growth of $t_{\rm QSS}$ with~$N$ could be interpreted as a breakdown of ergodicity, which is shown herein to occur for any value of~$\alpha$ in this interval.
\end{abstract}
%-----------------------------------------------------------------------
% \submitto{J. Stat. Mech.: Theory and Experiment}
% \noindent{\it Keywords\/}: Metastable states, Ergodicity breaking (Theory), Molecular dynamics
% \tableofcontents
  \maketitle
%-----------------------------------------------------------------------
\vspace{-0.20cm}
\section{Introduction}
\renewcommand{\baselinestretch}{1.15}
%-----------------------------------------------------------------------
Classical spin models have called the attention of statistical-mechanics 
and magnetism researchers throughout  
many years~\cite{Nakamura_1952Fisher_1964,Joyce_PR_1967,%
Stanley_Arbitrary_Dimensionality_PR_1969, stanleybook,cthompsonbook}. 
Many techniques have been used in their study, both analytical and numerical, in such a way that a reasonable knowledge of their
equilibrium thermodynamics has been achieved, and specially, of their critical properties. 
Among those models, one could mention the \mbox{$n$-vector} classical spin models, 
which present the XY ($n=2$) and Heisenberg ($n=3$) as particular cases.

An interesting formulation of a $n$-vector classical model comes 
when one adds a kinetic term to its Hamiltonian, i.e., the spin variables may be interpreted 
as classical rotators (see, e.g., Refs.~\cite{AntoniRuffo1995, AnteneodoTsallisPRL1998, NobreTsallisPRE2003}), 
so that the terminology ``inertial model'' is currently used.   
This additional term does not pose difficulties in the calculation of equilibrium properties within a 
canonical-ensemble approach, but it turns possible to derive equations
of motion for each rotator, which can be integrated by means of a molecular-dynamics procedure. In this way, the dynamical behaviour 
of these models can be investigated numerically without the need of introducing any particular type of 
probabilistic transition rules for changing the microscopic states. 

The inertial ferromagnetic XY model was introduced in Ref.~\cite{AntoniRuffo1995}, within a fully-coupled framework 
(i.e., infinite-range interactions), a limit where the mean-field approach becomes exact. 
Mostly referred to as ``Hamiltonian Mean Field'' (HMF) model, 
it became paradigmatic in the study of the dynamical behaviour of 
classical many-body Hamiltonian systems, and it has given rise to a large amount of 
works~\cite{LatoraRapisardaTsallisPRE2001, Yamaguchi_etalPA2004,%
PluchinoLatoraRapisardaPA2004PluchinoLatoraRapisardaPA2006PluchinoLatoraRapisardaPD2004, LatoraRapisardaRuffoPRL1998LatoraRapisardaRuffoPD1999,%
MoyanoAnteneodoPRE2006, PluchinoRapisardaTsallisEPL2007PluchinoRapisardaTsallisPA2008,%
ChavanisCampaEPJB2010CampaChavanisEPJB2013,
EttoumiFirpoPRE2013, CampaDauxoisRuffo_PR2009, RochaAmatoFigueiredoPRE2012}.
One of the most interesting features in the HMF model concerns the appearance 
of metastable states, for some particular initial conditions, whose
lifetime grows by increasing the total number of rotators, that
could be interpreted as a breakdown of ergodicity in the thermodynamic limit.    

A generalization of the HMF model was proposed in Ref.~\cite{AnteneodoTsallisPRL1998}, 
by introducing a distance~$r_{ij}$ between rotators at sites~$i$ and~$j$ of a given lattice.
Moreover, the two-body interaction was assumed to decay with~$r_{ij}$, like a power-law, 
$1/r_{ij}^{\alpha}$ ($\alpha \ge 0$). 
The index~$\alpha$ controls the range of the interactions, in such 
a way that one recovers both the HMF and nearest-neighbour-interaction
models in the particular limits  $\alpha=0$ and $\alpha\to\infty$, respectively.   
In between these two limits, one finds an important change of behaviour 
in the thermodynamic quantities, yielding two physically distinct regimes, 
namely, the long- ($\alpha \leq d$) and short-range ($\alpha > d$)
interaction regimes~\cite{AnteneodoTsallisPRL1998,JundKimTsallisPRB1995,%
CampaGiansantiMoroniTsallisPLA2001, TamaritAnteneodo_PRL_2000}.
In the latter, one has the usual extensive and intensive quantities, whereas in 
the former, one may find also nonextensive thermodynamic 
quantities~\cite{TsallisBook2009}.  
This generalization is usually referred to as~$\alpha$-XY 
or~$\alpha$-HMF model, and it also has been studied by several
groups~\cite{CampaDauxoisRuffo_PR2009,CampaGiansantiMoroniTsallisPLA2001,TamaritAnteneodo_PRL_2000,CampaGiansantiMoroni2000_PRE,% 
CirtoAssisTsallisPA2014,FirpoRuffoJPA2001VallejosAnteneodoPRE2001}.

The dynamics of the fully-coupled inertial version of the ferromagnetic 
Heisenberg model has been much less investigated in the 
literature~\cite{NobreTsallisPRE2003,NobreTsallisPA2004, GuptaMukamelPRE2013}.
Metastable, or quasi-stationary states (QSSs), also occur
for this system, similar to those that appear in the HMF model; 
their lifetime diverge by increasing the total number of rotators~$N$,
which implies that the order in which the thermodynamic limit 
(${N\to\infty}$) and the infinite-time limit (${t\to\infty}$) are
considered, becomes important.
More specifically, if we let ${N\to\infty}$ first, the system remains trapped in these
QSSs, never reaching the final Boltzmann-Gibbs (BG) equilibrium state,
most probably being the QSS itself the final state in such a case.
In these metastable states, thermodynamical quantities, like 
temperature and magnetization, do not coincide with the canonical-ensemble predictions.

Herein, we will modify the fully-coupled inertial ferromagnetic Heisenberg 
model, by introducing a distance~$r_{ij}$ between rotators at sites $i$ 
and~$j$ of a given lattice. Analogously to the $\alpha$-HMF 
model~\cite{AnteneodoTsallisPRL1998}, the rotator-rotator interaction will
be taken to decay with the distance like a power-law, 
$1/r_{ij}^{\alpha}$ ($\alpha \ge 0$). 
The interaction part of this Hamiltonian has already 
been studied analytically in Ref.~\cite{CampaGiansantiMoroniJPA2003},
within a canonical-ensemble approach to the equilibrium state of 
the corresponding \mbox{$d$-dimensional} model,   
where it was shown that if the two-body interaction is appropriately scaled,
a universal thermodynamical behaviour is achieved for $0 \leq \alpha < d$, 
e.g., relations involving temperature~$T$, magnetization~$M$, and
internal energy~$U$ become $\alpha$-independent in this interval.  
Apart from this study of the equilibrium state, an investigation 
of the dynamical behaviour of the above-mentioned Heisenberg  
model, and particularly, how the QSSs may be affected by the 
exponent~$\alpha$, has never been addressed in the literature, to our knowledge. 
In the present work we study this model on a ring, i.e, $d=1$ with
periodic boundary conditions, using  molecular-dynamics simulations.
We perform a detailed analysis of the QSSs behaviour by varying the energy, number of particles,
and range of the interaction, i.e., the exponent~$\alpha$. 
In the next section we define the model, the appropriate scaling 
for the interactions, the canonical-ensemble solution, equations 
of motion, and initial conditions to be used in the molecular-dynamics procedure.
In Section~\ref{Sec:Results} we present the results of our simulations, showing an 
agreement with some analytical results known in the literature
for the equilibrium state, in some particular limits; most 
importantly, we show the existence of QSSs for energies~$U$ below 
criticality ($U<U_{c}$). Such QSSs, which were verified herein for 
initial conditions corresponding to zero magnetization, 
appear for values of~$\alpha$ in the range $0 \leq \alpha <1$, being characterized by  
durations~$t_{\rm QSS}$ that increase for increasing values of~$N$, 
like those found numerically and analytically 
in previous works of the similar model of XY 
rotators~\cite{LatoraRapisardaTsallisPRE2001,Yamaguchi_etalPA2004, %
PluchinoLatoraRapisardaPA2004PluchinoLatoraRapisardaPA2006PluchinoLatoraRapisardaPD2004,LatoraRapisardaRuffoPRL1998LatoraRapisardaRuffoPD1999,%
MoyanoAnteneodoPRE2006, PluchinoRapisardaTsallisEPL2007PluchinoRapisardaTsallisPA2008,%
ChavanisCampaEPJB2010CampaChavanisEPJB2013,
EttoumiFirpoPRE2013, CampaDauxoisRuffo_PR2009, RochaAmatoFigueiredoPRE2012,%
CirtoAssisTsallisPA2014,CampaGiansantiMoroniPA2002,GiansantiMoroniCampaCSF2002}. 
Finally, in Section~\ref{Sec:Conclusions} we present our main conclusions. 
%-----------------------------------------------------------------------
\section{The model}  
%-----------------------------------------------------------------------
First, let us define the model in terms of~$N$
localized interacting classical rotators, on a 
$d$-dimensional hypercubic lattice, through the Hamiltonian
\begin{equation}
\label{eq:Hamiltonian_Heisenberg}
\mathcal{H}\,=\,\frac{1}{2}\sum_{i\,=\,1}^N \mathbf{L}_i^2 \,+\, \frac{1}{2\Ntil}\sum_{i\,=\,1}^N \sum_{\substack{ {j\,=\,1}\\{j\,\neq\,i} } }^N \frac{1- \mathbf{S}_i\cdot\mathbf{S}_j }{r_{ij}^\alpha}
\,=\, K \,+\, V~,
\end{equation}
% \vskip \baselineskip 
% \noindent
where $\alpha \ge 0$ and $\mathbf{S}_i$ represents a vector 
with~$n=3$ components, assigned to the rotator 
at site~$i$ (similar to a classical Heisenberg spin variable), 
allowed to change its direction continuously inside a 3-dimensional sphere 
of unity radius, leading to the constraint
$\mathbf{S}_i\cdot\mathbf{S}_i=S_i^{\,2}=1\,\,(\forall\,\,i)$~\footnote{This constraint reduces the number of degrees of freedom per particle from three to two.}.
Moreover, $\mathbf{L}_i$ depicts the corresponding 
angular momentum (we are considering moments of inertia equal to unit, so that
angular momenta and angular velocities are equivalent quantities).
The coupling constants may be identified as $J_{ij}=1/ (\Ntil r_{ij}^\alpha)$, where~$r_{ij}$ measures the (dimensionless) distance between rotators~$i$ and~$j$, 
defined as the minimal one, given that periodic boundary conditions will be considered.

One should notice that the interaction term defined 
in Eq.~(\ref{eq:Hamiltonian_Heisenberg}) is 
long-ranged for $\alpha \leq d$, in the sense
that if~$\Ntil \sim \mathcal{O}\de{1}$ 
the partition function does not admit a well-defined 
thermodynamic limit~\cite{CampaGiansantiMoroniJPA2003}, e.g., 
the internal energy per particle diverges in the limit~$N\to\infty$, 
so that the system is said to be nonextensive~\cite{TsallisBook2009}.
Consequently, in order to calculate the thermodynamic limit adequately  
whenever~$\alpha \leq d$, the quantity~$\Ntil$ has to be defined 
appropriately to ensure a total energy proportional to the system size~$N$.
For~$\alpha=0$ this is attained with~$\Ntil=N$, the so-called Kac's 
prescription~\cite{stanleybook,cthompsonbook,CampaDauxoisRuffo_PR2009,CampaGiansantiMoroniJPA2003}.
For a general $0\leq (\alpha/d) < \infty$, the extensivity of the 
Hamiltonian in~(\ref{eq:Hamiltonian_Heisenberg})
is achieved through the following choice for~$\Ntil$,
\begin{equation}
\label{eq:Ntil_Definicao}
{\widetilde N} \,\equiv\, \frac{1}{N}\sum_{i\,=\,1}^N\sum_{\substack{ {j\,=\,1}\\{j\, \neq\,i} } }^N 
\frac{1}{r_{ij}^\alpha}~. 
\end{equation}
% \vskip \baselineskip
% \noindent
This may be seen intuitively if we note that for~$N$ large,  
$\Ntil\sim N^{1-\alpha/d}$, if $0 \leq (\alpha/d) < 1$ 
(hence,~$\Ntil\sim N$ for $\alpha=0$), $\Ntil\sim \ln N$, if $(\alpha/d)=1$, 
and $\Ntil\sim 1/(\alpha/d -1)\sim\mathcal{O}\de{1}$, if $(\alpha/d) > 1$;  
therefore, this proposal yields the interaction term in the 
Hamiltonian~(\ref{eq:Hamiltonian_Heisenberg}) proportional to the 
system size~$N$, for all $\alpha \geq 0$.

The need of the scaling in Eq.~(\ref{eq:Ntil_Definicao}), 
for systems characterized by long-range interactions, began to be realized
within a generalized ferrofluid model~\cite{JundKimTsallisPRB1995} 
(see also~\cite{Tsallis_Fractals_1995}).
This proposal was applied successfully to a controllable-range interaction 
($\alpha$-dependent) Curie-Weiss model, where it was shown 
numerically that by considering such a scaling, 
the magnetization per particle follows a thermodynamical 
equation of state,~$M=M\de{T}$, that becomes 
independent of~$\alpha$ in the nonextensive 
regime~\cite{CannasTamarit_PRB_1996SampaioAlbuquerqueFortunatoPRB1997}.
Moreover, applications of this scaling for Lennard-Jones-like systems were 
carried in Ref.~\cite{GrigeraPLA1996CurilefTsallis_PLA1999}.
In addition to the aforementioned works, the correctness of the 
conjectural scaling of Eq.~(\ref{eq:Ntil_Definicao})
has been verified in several other systems, 
suggesting that it should be valid for a wide variety of systems with 
long-range interactions~\cite{TsallisBook2009}.

A systematic analysis of Eq.~\eqref{eq:Ntil_Definicao}
has been particularly undertaken for the $\alpha$-XY model~\cite{AnteneodoTsallisPRL1998}, whose 
Hamiltonian is identical to Eq.~\eqref{eq:Hamiltonian_Heisenberg}, 
but $n=2$, i.e., $\vS_i$ is treated as a two-dimensional 
classical rotator, allowed to change its direction continuously inside a circle of unit radius. 
For this system, it was shown both 
analytically~\cite{CampaGiansantiMoroni2000_PRE, CampaGiansantiMoroniJPA2003} and 
numerically~\cite{TamaritAnteneodo_PRL_2000, CampaGiansantiMoroni2000_PRE, CirtoAssisTsallisPA2014,GiansantiMoroniCampaCSF2002}
that, for $0 \leq \alpha < d$, the prescription~(\ref{eq:Ntil_Definicao})
associates to the $\alpha$-XY model the same thermodynamical behaviour 
as the one previously known for the case $\alpha=0$~\cite{AntoniRuffo1995}.
Furthermore, the work of Ref.~\cite{CampaGiansantiMoroniJPA2003} 
has extended such a result to a general $n$-dimensional~$\vS_i$ ($n=1,2,\ldots$) 
unit vector.

Although the model of Eq.~(\ref{eq:Hamiltonian_Heisenberg}) 
has been defined on a $d$-dimensional hypercubic lattice, from now on
we will restrict ourselves to a ring, i.e., a one-dimensional chain with
periodic boundary conditions.  
Based on its successful use for the above-mentioned systems, we
will consider the scaling of Eq.~(\ref{eq:Ntil_Definicao})
for the present problem as well.  
%-----------------------------------------------------------------------
\subsection{The Canonical-Ensemble Solution}
%-----------------------------------------------------------------------
Within a canonical-ensemble solution, the kinetic term~$K$ of the
Hamiltonian in Eq.~(\ref{eq:Hamiltonian_Heisenberg}) does not 
bring difficulties in the evaluation of average values.  
Hence, the nontrivial part for the calculation of equilibrium properties 
of the Hamiltonian~(\ref{eq:Hamiltonian_Heisenberg})
comes from the interaction term~$V$. 
This term may become troublesome to deal analytically for dimensions $d>1$,
and to our knowledge, it has been investigated analytically
only for $d=1$, i.e., the linear chain, in some particular
regimes of $\alpha$, namely, $\alpha < d$ and $\alpha\to\infty$.
The model on an open linear chain with nearest-neighbour interactions 
(limit $\alpha \to \infty$) was solved exactly in Ref.~\cite{Nakamura_1952Fisher_1964},
whose solution is equivalent to the one considering periodic boundary conditions, 
in the thermodynamic limit~\cite{Joyce_PR_1967} (see also Refs.~\cite{stanleybook,cthompsonbook, Parsons_PRB_1977}).
On the other hand, the fully-coupled limit ($\alpha=0$) 
corresponds to the case 
where the mean-field approach becomes exact, so that 
the model may be solved through a relatively easy calculation 
(see, e.g., Ref.~\cite{NobreTsallisPRE2003}). 
As mentioned above, using the scaling 
of Eq.~(\ref{eq:Ntil_Definicao})
the solution for $\alpha=0$ applies to all 
$0 \leq \alpha < d$~\cite{CampaGiansantiMoroniJPA2003}.

Therefore, considering the model of 
Hamiltonian~(\ref{eq:Hamiltonian_Heisenberg})
defined on a ring, i.e., a one-dimensional chain with periodic boundary conditions, the solutions described 
above may be summarized, yielding for the internal energy and 
magnetization per particle respectively (we work with units such that $k_{\rm B}=1$), 
%-----------------------------------------------------------------------
\begin{equation}
\label{eq:Solucao_U}
U\de{M,T} \,=\,\l\{
\begin{array}{ll}
T \,+\, \displaystyle{\frac{1}{2}}\comu{1-L^{2}\de{M/T} } &; \qquad \textrm{if}\hspace{0.3cm} 0 \leq \alpha <1~, \\ \\
T \,+\, \displaystyle{\frac{1}{2}}\comu{1-L\de{1/2T} }    &; \qquad \textrm{if}\hspace{0.3cm} \alpha \to \infty~, 
\end{array}\r.
\end{equation}

 \vskip \baselineskip
 \noindent

\begin{equation}
\label{eq:Solucao_M}
M\de{T} \,=\,\l\{
\begin{array}{ll}
M = L\de{M/T} &; \qquad \textrm{if}\hspace{0.3cm} 0 \leq \alpha < 1~, \\ \\
M = 0         &; \qquad \textrm{if}\hspace{0.3cm} \alpha\to\infty~, 
\end{array}\r.
\end{equation}
%-----------------------------------------------------------------------
where $M\de{T}$ denotes the modulus of the magnetization vector, 
\begin{equation}
\label{eq:Magnetizacao_Definicao}
\mathbf{M}\,=\,\frac{1}{N}\sum_{i\,=\,1}^N \mathbf{S}_i~. 
\end{equation}
In these equations, $L(x)$ represents the Langevin function,  
\[
L\de{x} \,=\, \coth x \,-\, \frac{1}{x} \,=\, \frac{I_{3/2}\de{x}}{I_{1/2}\de{x}}~, 
\]
and $I_n(x)$ is the modified Bessel function of the first kind.

\begin{figure}%[h!]
\centering
  \includegraphics[width=0.494\linewidth]{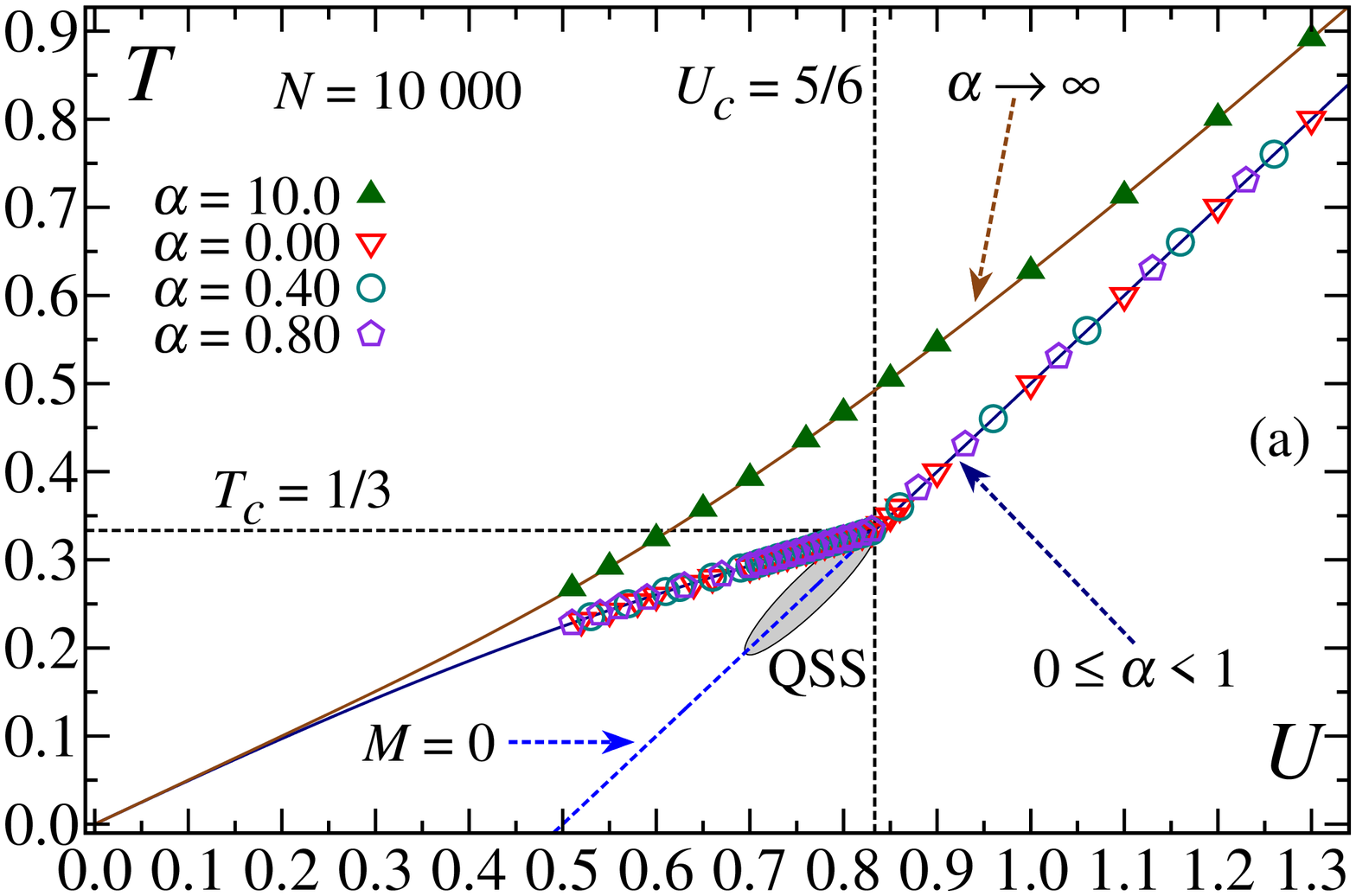}
  \includegraphics[width=0.494\linewidth]{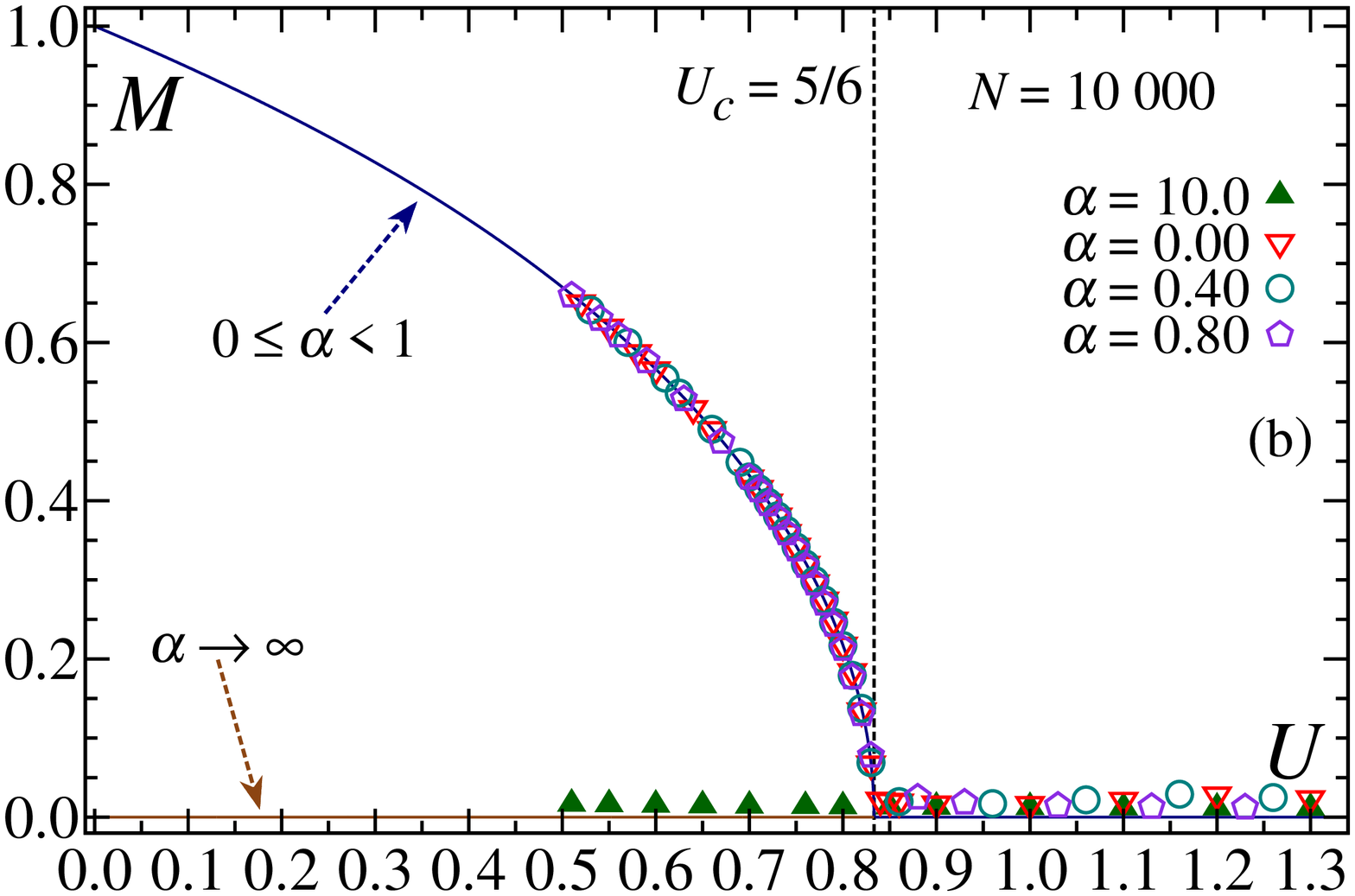}
\caption{\textsf{The good agreement between the analytical results [full lines, cf. Eqs.~(\ref{eq:Solucao_U}) and (\ref{eq:Solucao_M})] and data 
from numerical simulations of a system composed by $N=10\,000$ rotators,
in the long-time limit (symbols), is presented for the controllable-range
inertial Heisenberg model defined by Eq.~(\ref{eq:Hamiltonian_Heisenberg}).
(a) The kinetic temperature $T=\langle K(t) \rangle_{\textrm{time}} /N$
(which coincides with the temperature~$T$ at equilibrium) versus the
internal energy~$U$; (b) The magnetization~$M$ versus the internal energy~$U$. 
In panel (a) (in panel (b)) we compare numerical data, for typical
values of~$\alpha$, with the caloric curves (with the magnetization curve) 
in the cases $0 \leq \alpha < 1$~\cite{NobreTsallisPRE2003, CampaGiansantiMoroniJPA2003}, as well 
as in the limit $\alpha \to \infty$~\cite{Nakamura_1952Fisher_1964, Joyce_PR_1967}. 
In the former case, one has a second-order phase transition 
at the critical point $(T_{c}=1/3, \, U_{c}=5/6)$ (dashed lines), 
separating the ferromagnetic $(U<U_{c})$ and paramagnetic $(U>U_{c})$ phases.
The metastable solution corresponding to the quasi-stationary state (QSS) that appears for $U<U_{c}$,
characterized by zero magnetization (shaded region), and presenting a lower
kinetic temperature than the corresponding equilibrium solution, is also represented.}}
\label{Fig:Curvas_T_E_M}
\end{figure}

The analytical results presented above are displayed as solid lines 
in Fig.~\ref{Fig:Curvas_T_E_M}, where one finds a continuous phase transition
separating the ferromagnetic and paramagnetic states for values
of~$\alpha$ in the interval $0\leq \alpha < 1$; 
for~$\alpha\to\infty$, there is no phase transition at a finite temperature,
so that only the disordered phase with~$M=0$ exists.
In the first case, at sufficiently high temperatures 
(or equivalently, high enough energies), 
the directions of the~$\{ \mathbf{S}_i \}$ become randomly distributed, 
corresponding to the paramagnetic disordered phase, 
where one has the order parameter~$M=0$.
At the fundamental state all spins are parallel, 
corresponding to the ferromagnetically fully ordered case 
with~$M=1$. 
Hence, for all cases $0\leq \alpha < 1$, in between these two 
regimes, a continuous phase transition occurs at a critical temperature $T_{\textrm{c}}=1/3$, 
with an associated critical  energy $U_{\textrm{c}}=5/6\approx0.833$.
%-----------------------------------------------------------------------
%-----------------------------------------------------------------------

%-----------------------------------------------------------------------
\subsection{Equations of Motion}
%-----------------------------------------------------------------------
Herein we work with Cartesian components for the spin variables 
and angular momenta, written respectively as 
$\textbf{S}_i = \de{S_{x_i},S_{y_i},S_{z_i}}$ and   
$\textbf{L}_i = \de{L_{x_i},L_{y_i},L_{z_i}}$, yielding $6$
variables for each rotator. Hence, the set of $6N$ equations
to be solved numerically can be cast in the following 
form~\cite{Rapaport_Landau_PRE1996}, 
\begin{subequations}
% \begin{equation}
% \label{eq:Equacoes_Movimento}
%-----------------------------------------------------------------------
\begin{eqnarray}
\label{eq:Equacoes_MovimentoL}
\dot{\vL}_i &=& \vS_i \times \comu{ \frac{1}{\Ntil}\sum_{ \substack{ {j\,=\,1}\\{j\,\neq\, i}} }^N \frac{\vS_j}{r_{ij}^{\,\alpha}} }~,  
  \\  \nonumber \\
\label{eq:Equacoes_MovimentoS}
\dot{\vS}_i &=& \vL_i \times \vS_i~,
\end{eqnarray}
%-----------------------------------------------------------------------
\end{subequations}
% \end{equation}
% \vskip \baselineskip
% \noindent
for $i = 1,2,\ldots,N$. One notices that, in the particular case $\alpha=0$, the expression for~$\dot{\mathbf{L}}_i$ used in previous 
works~\cite{NobreTsallisPRE2003, NobreTsallisPA2004} is recovered, 
namely, $\dot{\mathbf{L}}_i = \mathbf{S}_i\times\mathbf{M}$ [where  
$\mathbf{M}$ represents the magnetization per particle of Eq.~(\ref{eq:Magnetizacao_Definicao})]. 

It is important to remind that  
% equations~(\ref{eq:Equacoes_MovimentoL})~and~(\ref{eq:Equacoes_MovimentoS}), 
Eqs.~\eqref{eq:Equacoes_MovimentoL} and~\eqref{eq:Equacoes_MovimentoS}, 
% equations~\eqref{eq:Equacoes_Movimento}, 
written in the Cartesian representation,   
are not canonical equations of motion, 
since $S_{\mu_i}$ and $L_{\mu_i}$ ($\mu=x,y,z$)
do not represent  canonically conjugate pairs.
Alternatively, one could also work along the line of Refs.~\cite{GuptaMukamelPRE2013, RobbReichlFaraggiPRE_2003},
by using spherical coordinates in order to write the spin variable as
$\mathbf{S}_i=\de{\cos\varphi_i\sin\theta_i, \sin\varphi_i\sin\theta_i, \cos\theta_i}$ 
and the squared angular momentum in the corresponding 
Lagrangian as $\mathbf{L}_i^2=L_{\theta_i}^2+L_{\varphi_i}^2/\sin^2\theta_i$. 
In this representation, one can derive equations of motion 
through the usual Hamiltonian formalism, where    
each rotator is characterized by two angles, 
$\theta_i\in[0,\pi)$ and $\varphi_i\in[0,2\pi)$,
and their canonically conjugate momenta, $L_{\theta_i}$ and $L_{\varphi_i}$.
Although the use of spherical coordinates, instead of the Cartesian ones  
of Eqs.~(\ref{eq:Equacoes_MovimentoL}) and (\ref{eq:Equacoes_MovimentoS}), 
seems to be a natural way to handle the problem, the denominator 
$\sin^{2}\theta_i$ that appears in~$\mathbf{L}^{2}_i$ becomes 
hard to be tackled numerically.
Indeed, as $\sin\theta_i$ approaches zero, one needs to decrease the 
time step used in the integrations of the equations of motion, in such 
a way that the computational time may increase substantially. 
This difficulty has been discussed in the literature by several 
authors~\cite{BarojasLevesqueQuentrecPRA1973,EvansMolPhys1977EvansMuradMolPhys1977,AllenTildesleyBook1987}, where
numerical techniques for circumventing it were proposed. 

Based on the above-mentioned reasons, herein we shall deal 
with the equations of motion
expressed in terms of Cartesian components, as in 
Eqs.~(\ref{eq:Equacoes_MovimentoL}) and (\ref{eq:Equacoes_MovimentoS}). 
It is straightforward to verify that, in addition to the total energy, 
the total angular momentum, $\vL=\sum_{i=1}^N \vL_i$, as well as 
the norm of each spin, $S_i=\comu{\vS_i \cdot \vS_i }^{1/2}$, are also 
integrals of motion.
Indeed, evaluating the time derivative of~$S_i$ and taking into account 
that~$\dot{\vS}_i \perp \vS_i$, we get
\[
\frac{\rd S_i}{\rd t}
% = \frac{\rd }{\rd t} \comu{\mathbf{S}_i \cdot \mathbf{S}_i }^{1/2} 
\,=\, \frac{\dot{\mathbf{S}}_i \cdot \vS_i}{S_i} \,=\, 0~. 
\]
% \vskip \baselineskip
% \noindent
Similarly for $\dot{\mathbf{L}}$, 
\[
% \dfrac{\rd\mathbf{L} }{\rd t} = 
\sum_{i\,=\,1}^N \dot{\vL}_i \,=\, \de{\sum_{i\,=\,1}^N \mathbf{S}_i} \times \de{\frac{1}{\Ntil} \sum_{j\,=\,1}^N \frac{\mathbf{S}_j}{r_{ij}^{\alpha}}} \,=\, 0~. 
\]
% \vskip \baselineskip
% \noindent
It should be stressed that the equation of motion~(\ref{eq:Equacoes_MovimentoL})  
corresponds precisely to the Euler equation for a linear rigid body of unit length 
(and unit inertial moments).
In such system the angular momentum~$\vL_i$ is always perpendicular 
to~$\vS_i$ (the ``molecular" axis), yielding the constraint 
$\mathbf{L}_i\cdot\mathbf{S}_i=0$, 
to be incorporated in the initial state.
Once this additional constraint is imposed, it will hold throughout the whole time 
evolution, since the product $\mathbf{L}_i\cdot\mathbf{S}_i$ is also a constant 
of motion, as we can verify by 
using Eqs.~(\ref{eq:Equacoes_MovimentoL}) and (\ref{eq:Equacoes_MovimentoS}),   
\begin{equation}
\label{eq:ddtlisi}
\frac{\rd }{\rd t} \de{\vL_i\cdot\vS_i} \,=\, \dot{\vL}_i\cdot\vS_i \,+\, \vL_i\cdot\dot{\vS}_i \,=\, 0~. 
\end{equation}
%-----------------------------------------------------------------------
\subsection{\label{Sec:Numerical_Procedure}Numerical Procedure and Initial Conditions}
%-----------------------------------------------------------------------
All our molecular-dynamical simulations were carried for a 
single copy of the system defined 
by Eq.~(\ref{eq:Hamiltonian_Heisenberg}),  
considering fixed values of the 
total number of rotators~$N$ and energy per particle~$U$. 
Since we are applying periodic boundary conditions,
the model of Eq.~(\ref{eq:Hamiltonian_Heisenberg})
becomes defined on a ring, with~$r_{ij}$ corresponding to the minimal
dimensionless distance between 
rotators~$i$ and~$j$, taking the values $1, 2, 3\ldots N/2$, for each rotator~$i$.  
 To integrate the~$6N$ equations of motion 
[Eqs.~(\ref{eq:Equacoes_MovimentoL}) and (\ref{eq:Equacoes_MovimentoS})]
we have used a standard fourth-order Runge-Kutta scheme, 
with an integration step chosen in such a way to yield conservation 
of the energy per particle within a relative fluctuation 
smaller than $10^{-5}$; this was achieved with a time step
 typically\footnote{A substantial gain in computational time 
was obtained, in the simulations with $\alpha\neq 0$ and larger values of~$N$, by using a Fast-Fourier-Transform algorithm.
To implement this technique we have used the FFTW library \url{http://www.fftw.org}.} of~$\delta t = 0.02$.

The initial conditions used were such that each
angular momentum component~$L_{\mu_i}$ ($\mu=x,y,z$) was drawn 
at random from a uniform distribution, and then rescaled 
to yield zero total angular momentum for the whole system.
In what concerns the variables $S_{\mu_i}$, the initial conditions
corresponded to those of zero magnetization, 
achieved numerically as~$M \approx 0$ (typically $\sim10^{-3}$). 
For this, the components $S_{x_i}$ and $S_{y_i}$ were also drawn 
at random from a uniform distribution within the interval~$\comu{-1,1}$, 
whereas the components $S_{z_i}$ were set in  
such way to preserve the constraint~$\vL_i\cdot\vS_i=0$ 
$(\forall i)$, i.e., through\footnote{It should 
be mentioned that this constraint was not used in Refs.~\cite{NobreTsallisPRE2003, NobreTsallisPA2004},  
so that $\vL_i\cdot\vS_i=\delta_{i}$ $(\delta_{i} \in [-1,1])$ in these works,
at the beginning of the simulations. However, it was  verified numerically that $\delta_{i}$ was also a constant 
of motion, in such a way that Eq.~(\ref{eq:ddtlisi}) remained true $(\forall i)$.}
$S_{z_i} = -\de{ L_{x_i}S_{x_i} + L_{y_i}S_{y_i} }/L_{z_i}$.
This procedure leads to unnormalized spins, which are then normalized by dividing each component by 
$\sqrt{S_{x_i}^2+S_{y_i}^2+S_{z_i}^2}$~; finally, all 
$L_{\mu_i}$'s were rescaled again to obtain precisely
the desired total energy~$U$. Consistently with  Eqs.~(\ref{eq:Solucao_U}) and (\ref{eq:Solucao_M}),
initial configurations with zero magnetization like the one 
just described above, only applies to a total energy $U>1/2$.
%-----------------------------------------------------------------------
\section{\label{Sec:Results}Results}
%-----------------------------------------------------------------------
Although we are dealing with three Cartesian components for the angular momenta ($L_{x_i}, L_{y_i}$ and  $L_{z_i}$), due to 
the constraint~$\vL_i\cdot\vS_i=0$ only two of these components are independent. Let us then define the instantaneous kinetic temperature 
as given by $T\de{t} = K\de{t}/{N}$. 
Since we are working with a single copy of the system defined by Eq.~(\ref{eq:Hamiltonian_Heisenberg}), the 
time average at the equilibrium state should be identified with 
the thermodynamic temperature, $T = \media{T\de{t}}_{\textrm{time}}$, and in a similar way, one has for the equilibrium 
magnetization, $M = \media{M\de{t}}_{\textrm{time}}$. 
As it will be shown later on, such time averages should be computed after a sufficiently 
long time, where both~$T\de{t}$ and $M\de{t}$ fluctuate around the 
values predicted by Boltzmann-Gibbs (BG) statistical mechanics.
In this regime, where ergodicity is expected to hold, 
time averages and ensemble averages should coincide.  
 
In figure~\ref{Fig:Curvas_T_E_M} we present results from 
analytical calculations (full lines) and data from numerical simulations in the long-time limit (symbols), 
for the controllable-range inertial Heisenberg model of Eq.~(\ref{eq:Hamiltonian_Heisenberg}).
In figure~\ref{Fig:Curvas_T_E_M}(a) we represent the kinetic temperature
versus the internal energy~$U$, whereas in Fig.~\ref{Fig:Curvas_T_E_M}(b) we do the same for the magnetization.
As mentioned above, at equilibrium, an average over the quantity~$T\de{t}$ is expected to coincide with the temperature~$T$
(obtained from the equipartition theorem), so that for the analytical results presented, the vertical axis
of Fig.~\ref{Fig:Curvas_T_E_M}(a) corresponds precisely to the equilibrium temperature~$T$, 
whereas for the numerical ones, it represents  the quantity computed from a time average. 
All data shown in Fig.~\ref{Fig:Curvas_T_E_M}
for both temperature and magnetization  coincide with the analytical results, showing 
that the equilibrium state considered numerically should be the one predicted by BG theory.  
The canonical-ensemble solution of the case $\alpha=0$ (see, e.g., Ref.~\cite{NobreTsallisPRE2003}) is represented by the full line
in Fig.~\ref{Fig:Curvas_T_E_M}(a) (usually known as the caloric curve),   
presenting a discontinuity in its slope at the critical point $(T_{c}=1/3, U_{c}=5/6)$, where the magnetization becomes zero.
According to Ref.~\cite{CampaGiansantiMoroniJPA2003}, the equilibrium 
analytical solution of the case $\alpha=0$ should apply to any $0 \leq \alpha < 1$, 
so that the corresponding full lines in  Figs.~\ref{Fig:Curvas_T_E_M}(a) and~(b) hold for any~$\alpha$ in this range. 
Such a universal behaviour was verified numerically for 
the $\alpha$-XY model~\cite{TamaritAnteneodo_PRL_2000,%
CampaGiansantiMoroni2000_PRE,CirtoAssisTsallisPA2014,GiansantiMoroniCampaCSF2002}, 
and it is now established herein for the Heisenberg model, as displayed in 
Fig.~\ref{Fig:Curvas_T_E_M} for typical values of $0 \leq \alpha < 1$. 
For completeness, in Fig.~\ref{Fig:Curvas_T_E_M}(a) 
we also present the solution corresponding to the quasi-stationary state~(QSS) hat appears for $U<U_{c}$, 
characterized by zero magnetization, leading to $T=U-1/2$,  
and presenting a lower kinetic temperature than the corresponding equilibrium solution. 
It was essentially within the energy range $0.70 \lesssim U < U_c = 5/6$
that QSSs have been investigated within the present numerical investigation. 
In the limit $\alpha \to \infty$ our model recovers the
nearest-neighbour-interaction inertial Heisenberg model on a
linear chain, for which the interaction part was solved exactly in Refs.~\cite{Nakamura_1952Fisher_1964,Joyce_PR_1967}, and shown to 
present  zero magnetization for all $T>0$; consequently, 
its caloric curve exhibits the smooth behaviour indicated in Fig.~\ref{Fig:Curvas_T_E_M}(a). 
In this case, our simulations for~$\alpha=10$ are in full agreement with the results of the analytical solutions. 
%-----------------------------------------------------------------------
\begin{figure}%[h!]
  \centering
  \includegraphics[width=0.75\linewidth]{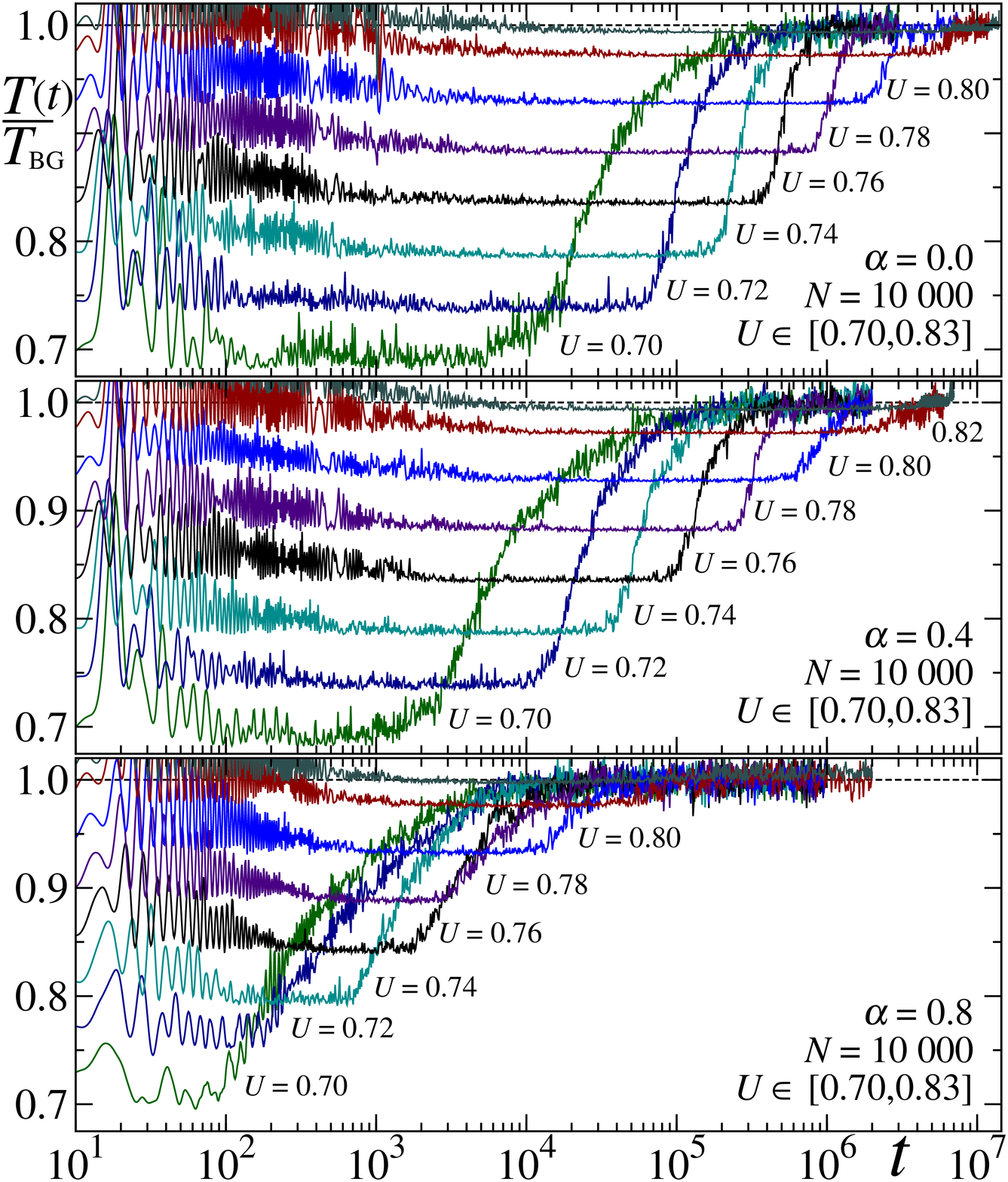}
\caption{\textsf{Time evolution of the dimensionless quantity $T(t)/T_{\rm BG}$ 
for typical values of the total energy~$U$ 
($U<U_{\textrm{c}}$), for a single copy 
of $N=10\,000$ Heisenberg rotators, in the cases 
$\alpha=0.0,0.4$, and $0.8$ (from top to bottom). 
For each value of~$U$, the corresponding equilibrium
temperature $T_{\rm BG}$ is obtained from the caloric curve in 
Fig.~\ref{Fig:Curvas_T_E_M}. 
The time is also dimensionless and each unit of (physical) time~$t$
corresponds to 50 iterations of the equations of motion.}}
\label{Fig:T_Vs_Tempo_Varios_U}
\end{figure}
%-----------------------------------------------------------------------

In figure~\ref{Fig:T_Vs_Tempo_Varios_U} we exhibit the time evolution of~$T(t)$ (conveniently rescaled in each case by the corresponding equilibrium temperature~$T_{\rm BG}$)
for typical values of energies, $U<U_{\textrm{c}}=5/6 \approx 0.833$, and 
$\alpha=0.0,0.4$, and $0.8$. 
Simulations were carried for a single copy of $N=10\,000$ rotators, considering the above-defined
initial conditions, for sufficiently long times (up to times slightly larger 
than $t=10^{7}$ in some cases). 
For all energies investigated, in the range $0.70 \leq U \leq 0.83$, 
we have verified the existence of QSSs and after
these, a crossover is observed to a state whose temperature and magnetization
coincide with those obtained analytically within BG statistical 
mechanics. For a given value of~$\alpha$ one has that: 
(i) The lower energies have produced QSSs with smaller 
durations~$t_{\rm QSS}$, and one observes that~$t_{\rm QSS}$ 
increases substantially as one approaches the critical energy; 
(ii) The gaps separating~$T(t)$ of these  QSSs and their corresponding values of equilibrium
temperatures~$T_{\rm BG}$ are larger for smaller energies, 
dropping to zero as $U \rightarrow U_{c}$. 
On the other hand, for a fixed total energy, higher values of~$\alpha$
yield smaller~$t_{\rm QSS}$, and it will be shown later on
that these durations tend to zero as $\alpha \to 1$.   
Therefore, the QSSs disappear in both limits $U \rightarrow U_{c}$ and $\alpha \to 1$. 
The existence of QSSs for energies below, but sufficiently close 
to~$U_{c}$, is well known to occur in the HMF and $\alpha$-XY (with $0 < \alpha <1$) models, and it has been found analytically and numerically by many 
authors~\cite{LatoraRapisardaTsallisPRE2001,Yamaguchi_etalPA2004, %
PluchinoLatoraRapisardaPA2004PluchinoLatoraRapisardaPA2006PluchinoLatoraRapisardaPD2004,LatoraRapisardaRuffoPRL1998LatoraRapisardaRuffoPD1999,%
MoyanoAnteneodoPRE2006,%
PluchinoRapisardaTsallisEPL2007PluchinoRapisardaTsallisPA2008,%
ChavanisCampaEPJB2010CampaChavanisEPJB2013,%
EttoumiFirpoPRE2013, CampaDauxoisRuffo_PR2009,RochaAmatoFigueiredoPRE2012,CirtoAssisTsallisPA2014,CampaGiansantiMoroniPA2002,GiansantiMoroniCampaCSF2002}.
Moreover, the most interesting characteristic of such states concerns an increase in their lifetime
with the total number of rotators~$N$; this property will also be verified for the present model. 
%-----------------------------------------------------------------------
\begin{figure}%[h!]
 \centering
 \includegraphics[width=0.83\linewidth]{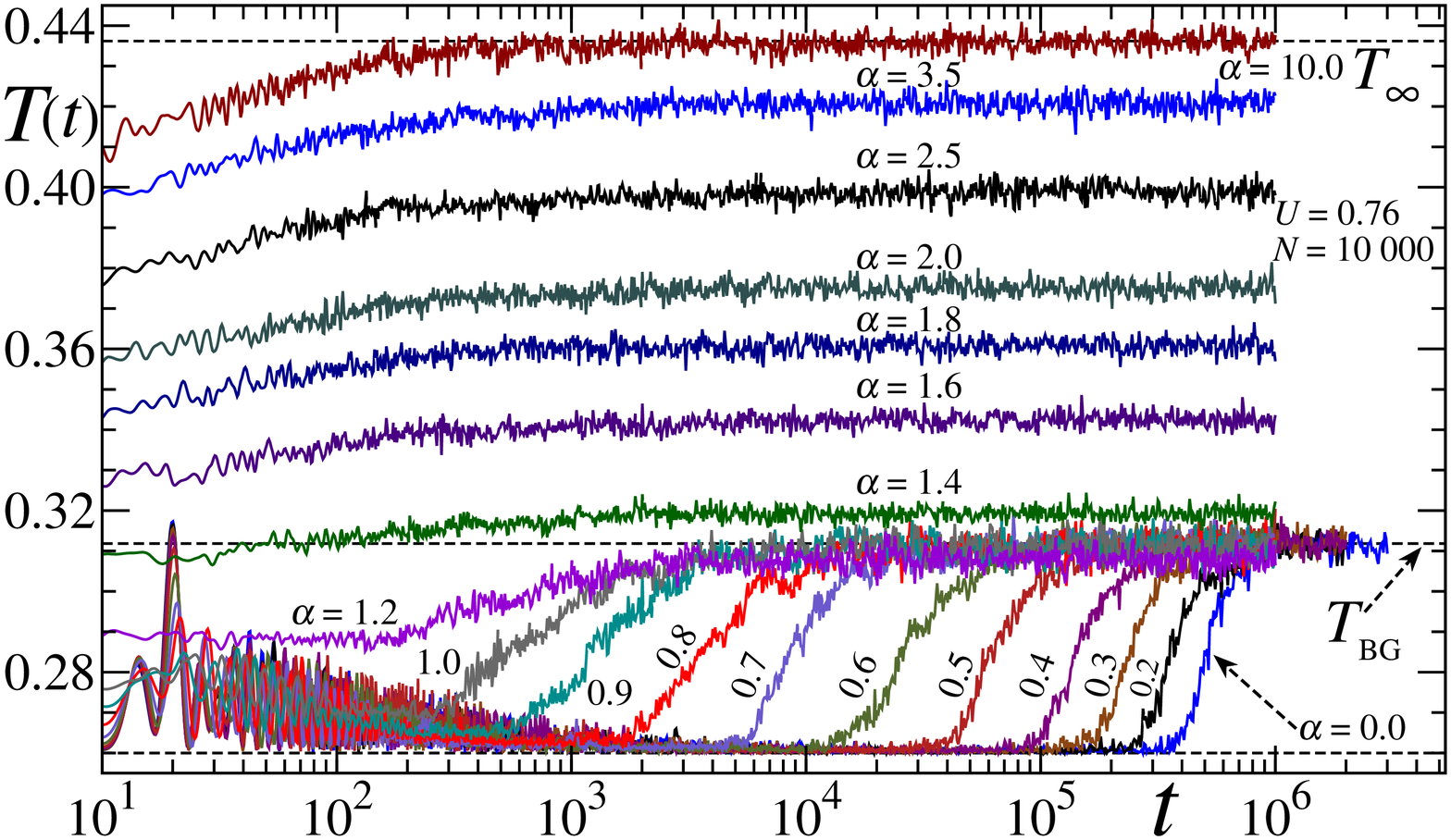}
\caption{\textsf{Time evolution of the kinetic temperature $T(t) = K(t)/N$ for a single realization of $N=10\,000$ Heisenberg rotators defined 
by Eq.~(\ref{eq:Hamiltonian_Heisenberg}), with  a total energy $U=0.76$, and several values of the 
interaction-range parameter~$\alpha$. 
The upper horizontal dashed line, at~$T_\infty\approx0.4362$, represents the Boltzmann-Gibbs~(BG) equilibrium temperature of 
the corresponding~\,$\alpha \to \infty$ model [cf. Eq.~(\ref{eq:Solucao_U})].
The dashed horizontal line at~$T_{\rm BG}\approx0.3118$
indicates the BG equilibrium temperature for $0 \le \alpha < 1$~[see Eqs.~(\ref{eq:Solucao_U}) and (\ref{eq:Solucao_M})], 
whereas the lower horizontal line, at $T=U-1/2=0.26$, corresponds to 
the QSS temperature characterized by zero magnetization.
In the range $1\leq \alpha < \infty$, no analytical solution is available, 
as far as we know. The time is dimensionless and each unit of (physical)
time~$t$ corresponds to 50 iterations of the equations of motion.}}
\label{Fig:T_Vs_Tempo_Varios_Alphas}
\end{figure}
%-----------------------------------------------------------------------

The QSSs shown in Fig.~\ref{Fig:T_Vs_Tempo_Varios_U} 
appear throughout the whole range $0 \leq \alpha <1$, 
as exhibited 
in Fig.~\ref{Fig:T_Vs_Tempo_Varios_Alphas} for several values of~$\alpha$.
Each case corresponds to molecular-dynamics simulations of 
a single realization of $N=10\,000$ Heisenberg rotators defined 
by Eq.~(\ref{eq:Hamiltonian_Heisenberg}), with  
a total energy $U=0.76$. One sees that 
the lifetime of these QSSs decrease for increasing values
of~$\alpha$ in the range $0 \leq \alpha <1$
and that all QSSs occur at a kinetic temperature 
corresponding to zero magnetization (or slightly larger than zero),  
i.e., $T \gtrsim U-1/2=0.26$.  
After these QSSs, the system approaches the BG equilibrium 
temperature ($T_{\rm BG}\approx0.3118$), associated to 
its energy through the caloric curve of 
Fig.~\ref{Fig:Curvas_T_E_M}. 
At $\alpha \approx 1$ one notices a crossover
from the existence of these metastable states 
to a smooth gradual increase in the kinetic temperature, such
that for $\alpha>1$ there is no clear evidence of QSSs. 
Finally, for a sufficiently 
high value of~$\alpha$ (e.g., $\alpha=10$), the long-time behaviour 
of the kinetic temperature~$T(t)$ gradually approaches the BG equilibrium temperature of
the corresponding~\,$\alpha \to \infty$ model, i.e.,  
the nearest-neighbour-interaction model ($T_\infty\approx0.4362$).
Our numerical results however do not exclude a  
possible finite-size dependence of these short-lived QSSs 
for~$\alpha \gtrsim 1$, as already found in the~$\alpha$-XY 
case~\cite{TurchiFanelliLeoncini2011EttoumiFirpoPRE2011}. 
However, no dependence on $N$ is expected for $\alpha\gg 1$,
since the interaction potential becomes effectively short range 
in this limit.
 
%-----------------------------------------------------------------------
\begin{figure}%[h!]
 \centering
 \includegraphics[width=0.65\linewidth]{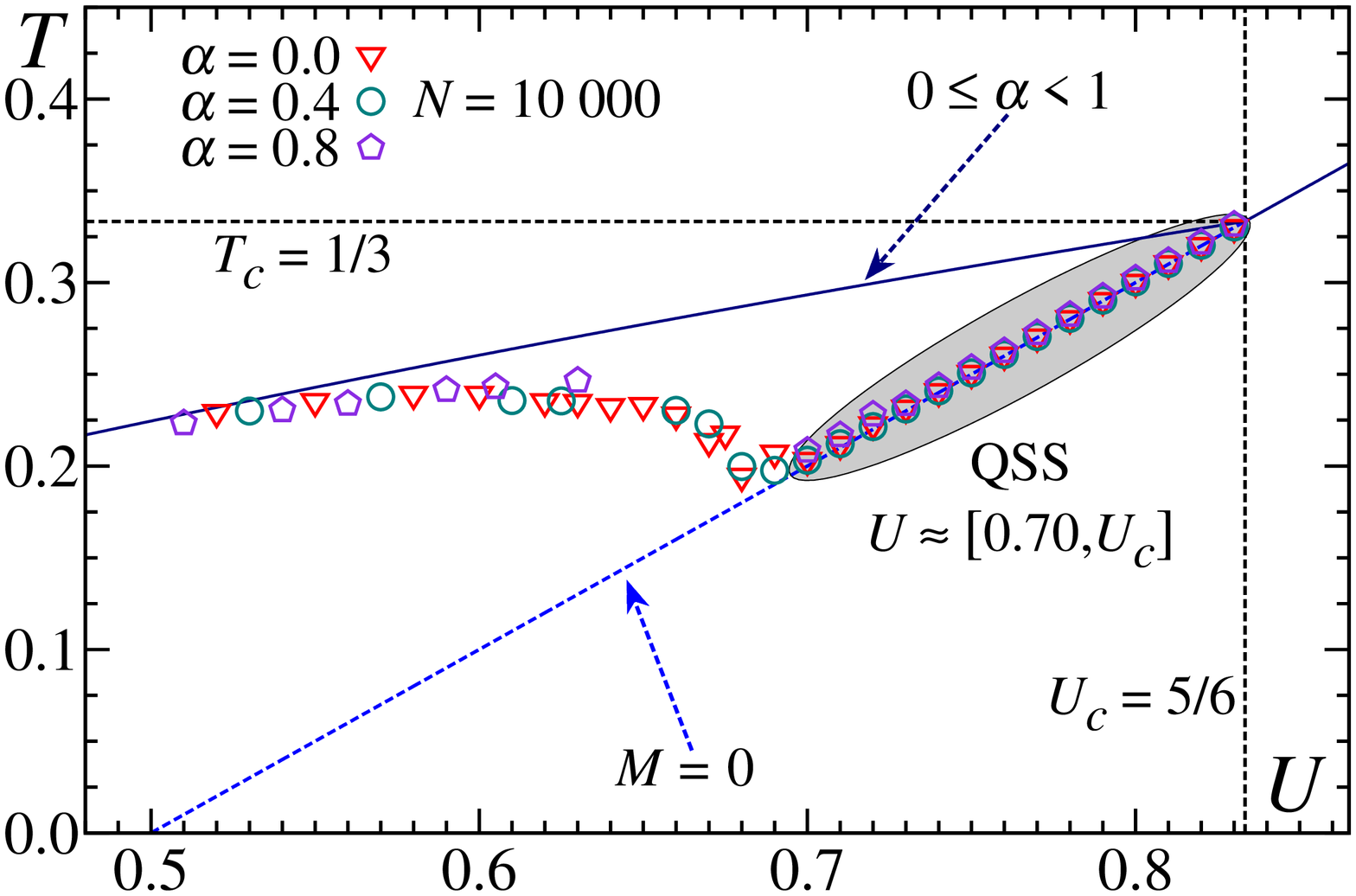}
\caption{\textsf{Part of the caloric curve for the cases $0 \leq \alpha <1$ 
[cf. Fig.~\ref{Fig:Curvas_T_E_M}(a)] is exhibited for energies $U \leq U_{c}$ (full line). The range of energies over which QSSs
were verified numerically is shown for both zero magnetization (dashed line), as well as for $M>0$. The symbols correspond to 
data associated with the kinetic temperatures of the QSSs, obtained from numerical simulations for a single copy of a system composed 
by $N=10\,000$ rotators, in the cases $\alpha=0.0,0.4$, and $0.8$.}}
\label{Fig:Curva_T_U_QSS}
\end{figure}
%-----------------------------------------------------------------------

An important question concerns the range of energies over which
the above-mentioned QSSs exist.  
In figure~\ref{Fig:Curva_T_U_QSS} we present 
part of the caloric curve for $0 \leq \alpha <1$, corresponding
to energies $U \leq U_{c}$ (full line). Numerical simulations for a single
copy of $N=10\,000$ rotators, in the cases $\alpha=0.0,0.4$, and $0.8$, 
show that the QSSs characterized by $M=0$
(symbols along the dashed line) exist within the 
range $0.70 \lesssim U < U_{c}$~, in agreement with the results of 
Fig.~\ref{Fig:T_Vs_Tempo_Varios_U}.
However, as one goes further below the critical point, roughly
for $0.50<U<0.70$,
QSSs are still found numerically, although characterized by finite values of  
magnetization, in spite of the initial conditions of zero magnetization considered.   
Such states are understood due to the prevalence of the ferromagnetic 
couplings for sufficiently low values of~$U$, and they may be observed 
more clearly in the case $\alpha=0.0$, being sometimes difficult to be identified
for higher values of~$\alpha$ (e.g., $\alpha=0.80$).
It should be noticed that within a small energy range 
(typically $U \lesssim 0.70$), the QSSs yield a 
negative microcanonical specific heat. Although this
represents a well-known feature for the HMF model
(see discussion in Ref.~\cite{ChavanisCampaEPJB2010CampaChavanisEPJB2013}), it 
is the first time that such a result is verified for a system
of Heisenberg rotators, to our knowledge. The numerical data 
of Fig.~\ref{Fig:Curva_T_U_QSS} suggests that a 
negative specific heat should occur for $0 \leq \alpha < 1$.   
%-----------------------------------------------------------------------
\begin{figure}%[h!]
 \centering
\includegraphics[width=0.75\linewidth]{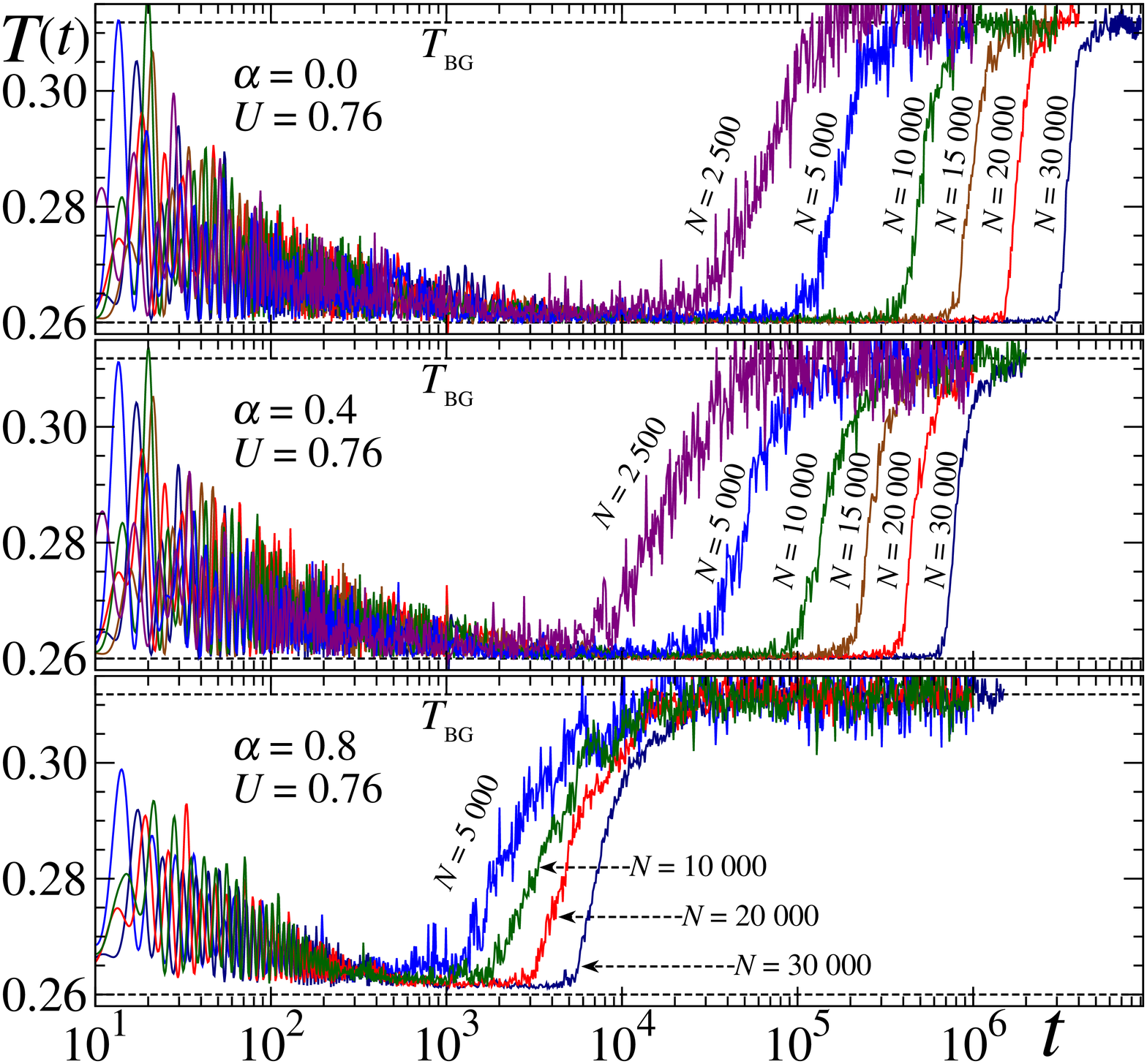}
\caption{\textsf{Typical QSSs presented in Figs.~\ref{Fig:T_Vs_Tempo_Varios_U} and~\ref{Fig:T_Vs_Tempo_Varios_Alphas} 
are analysed for varying system sizes, showing an increase in their 
lifetime~$t_{\rm QSS}$ as~$N$ increases.}}
\label{Fig:T_Vs_Tempo_3_Alphas}
\end{figure}
%-----------------------------------------------------------------------

The most interesting feature of the QSSs
presented in Figs.~\ref{Fig:T_Vs_Tempo_Varios_U}
and~\ref{Fig:T_Vs_Tempo_Varios_Alphas} concerns an increase in~$t_{\rm QSS}$ for  
increasing~$N$, as shown in Fig.~\ref{Fig:T_Vs_Tempo_3_Alphas}, 
where we analyse the QSSs for an energy~$U=0.76$ in 
the cases $\alpha=0.0,0.4$, and~$0.8$, by varying the system sizes. 
It should be mentioned that, in order to identify QSSs clearly
within the present approach, we had to simulate
systems with sufficiently high values of~$N$ (essentially, $N=2\,500$ or higher).
Hence, in Fig.~\ref{Fig:T_Vs_Tempo_3_Alphas}
such a behaviour is presented for typical values 
of~$N$ in the range from $N=2\,500$ up to $N=30\,000$, 
for the smaller values of~$\alpha$ (i.e., $\alpha=0.0$ and~$0.4$);
however, in the case $\alpha=0.8$ the observation of QSSs becomes
more difficult, in such a way that one needs to simulate
larger values of~$N$ (e.g., $N=5\,000$ or higher)
for this purpose. The growth of~$t_{\rm QSS}$ for increasing~$N$ represents 
a well-known feature in models of rotators, like the 
HMF and $\alpha$-XY (with $0 < \alpha <1$) 
models~\cite{LatoraRapisardaTsallisPRE2001,Yamaguchi_etalPA2004,%
PluchinoLatoraRapisardaPA2004PluchinoLatoraRapisardaPA2006PluchinoLatoraRapisardaPD2004,%
LatoraRapisardaRuffoPRL1998LatoraRapisardaRuffoPD1999,MoyanoAnteneodoPRE2006,PluchinoRapisardaTsallisEPL2007PluchinoRapisardaTsallisPA2008,%
ChavanisCampaEPJB2010CampaChavanisEPJB2013,%
EttoumiFirpoPRE2013, CampaDauxoisRuffo_PR2009,CirtoAssisTsallisPA2014,CampaGiansantiMoroniPA2002,GiansantiMoroniCampaCSF2002},
as well as the inertial Heisenberg model~\cite{NobreTsallisPRE2003,NobreTsallisPA2004, GuptaMukamelPRE2013},
yielding significant physical consequences, some of them directly
related to the applicability of BG statistical mechanics. 
More specifically, the order in which the thermodynamic limit (${N\to\infty}$) and the infinite-time limit (${t\to\infty}$) are
considered becomes mostly relevant in the present case, 
in such a way that if the thermodynamic limit is carried first, one should 
remain in a~QSS forever. 
This may be viewed as a breakdown of ergodicity, 
with the system being trapped in~a~QSS,    
never reaching the~BG equilibrium state; 
consequently, time averages and ensemble averages do not coincide, violating
a basic assumption of BG statistical mechanics.
The QSS represents the final state in this case;
whether it should be interpreted as a truly equilibrium
thermodynamical one represents an important question beyond the scope of the present work.  

\begin{figure}%[t!]
 \centering
 \includegraphics[width=0.65\linewidth]{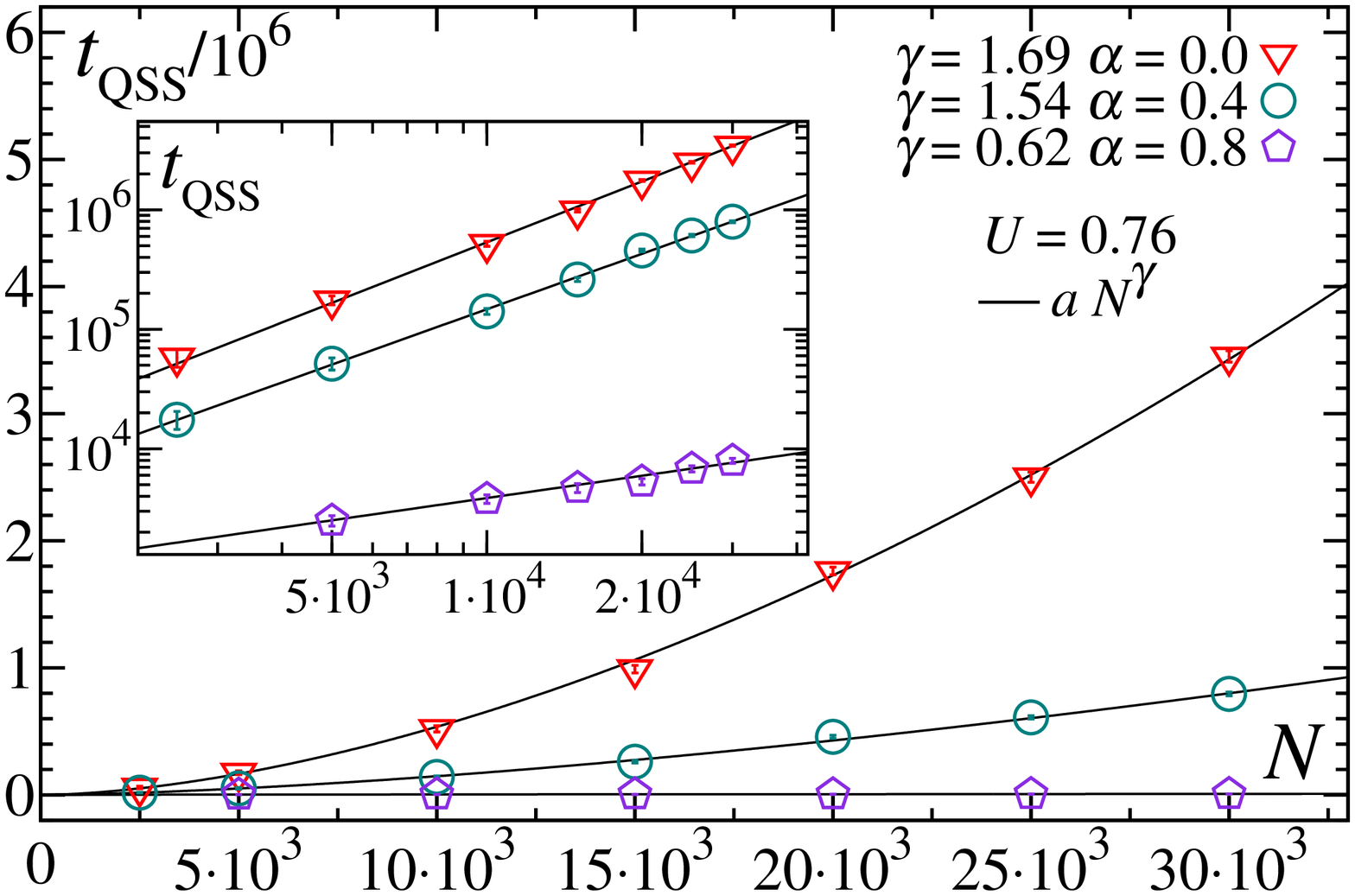}
\caption{\textsf{The growth of the lifetime of the QSSs ($t_{\rm QSS}$)
with respect to the size of the system is presented, for a fixed 
energy ($U=0.76$) and three typical values of~$\alpha$ ($\alpha=0.0,0.4$, and $0.8$).  
Symbols correspond to data from numerical simulations, whereas full lines 
stand for the fits proposed. In the inset we exhibit the same data in a log-log 
representation.}}
\label{Fig:Behavior_QSS_N}
\end{figure}

Herein we have defined the duration~$t_{\rm QSS}$ of a given 
QSS (e.g., those exhibited in Fig.~\ref{Fig:T_Vs_Tempo_3_Alphas}), as the time
at which~$T(t)$ presents its halfway between the values at the QSS and its corresponding~$T_{\rm BG}$. 
The fact that we are computing this time interval 
starting from $t=0$, including a short transient regime
before reaching the QSS, does not affect our final results,
since this small transient becomes negligible in the final 
computation of~$t_{\rm QSS}$, for sufficiently large
values of~$N$. According to this, the   
growth of~$t_{\rm QSS}$ with the total number of 
rotators~$N$, for a given energy ($U=0.76$) and three 
typical values of~$\alpha$ ($\alpha=0.0,0.4$, and $0.8$),
is exhibited in Fig.~\ref{Fig:Behavior_QSS_N}.   
In the three cases investigated, these behaviours are fitted
by power laws,  $t_{\rm QSS} \sim N^{\gamma}$
(as shown in the respective inset, where the same data 
is presented in a log-log representation), 
with the exponent~$\gamma$
decreasing as~$\alpha$ increases, e.g.,    
$\gamma \approx 1.69 \ (\alpha=0.0)$, 
$\gamma \approx 1.54 \ (\alpha=0.4)$, and
$\gamma \approx 0.62 \ (\alpha=0.8)$.
In the first case, our estimate agrees with the
one of Ref.~\cite{GuptaMukamelPRE2013} (see also~\cite{guptamukameljstat2011}), where $\gamma \approx 1.70$ was computed.
Curiously, such an estimate has also been found numerically for QSSs of the HMF 
model, with similar initial conditions
(zero magnetization)~\cite{Yamaguchi_etalPA2004}, being supported 
through analytical approaches in Ref.~\cite{EttoumiFirpoPRE2013}.
One should mention that in the case $\alpha=0.8$ 
the QSSs are more difficult to be observed, 
appearing only for sufficiently high values of~$N$, such
that data points for smaller~$N$ are not exhibited. 
In spite of such difficulties, our simulations suggest that 
one should have $\gamma \to 0$ in the limit $\alpha \to 1$. 
Although relevant, it is not  possible to propose a particular 
scaling behaviour describing the decrease of $\gamma$ with respect to 
$\alpha$, on the basis of the present numerical simulations. 
Such a scaling deserves further computational efforts, from
which one could obtain a larger number of data points for an appropriate
plot of $\gamma$ versus $\alpha$ (see, e.g., Ref.~\cite{BachelardKastnerPRL2013} for the XY case).

%-----------------------------------------------------------------------
%-----------------------------------------------------------------------

%-----------------------------------------------------------------------
\begin{figure}[t!]
\centering
\includegraphics[height=0.325\linewidth]{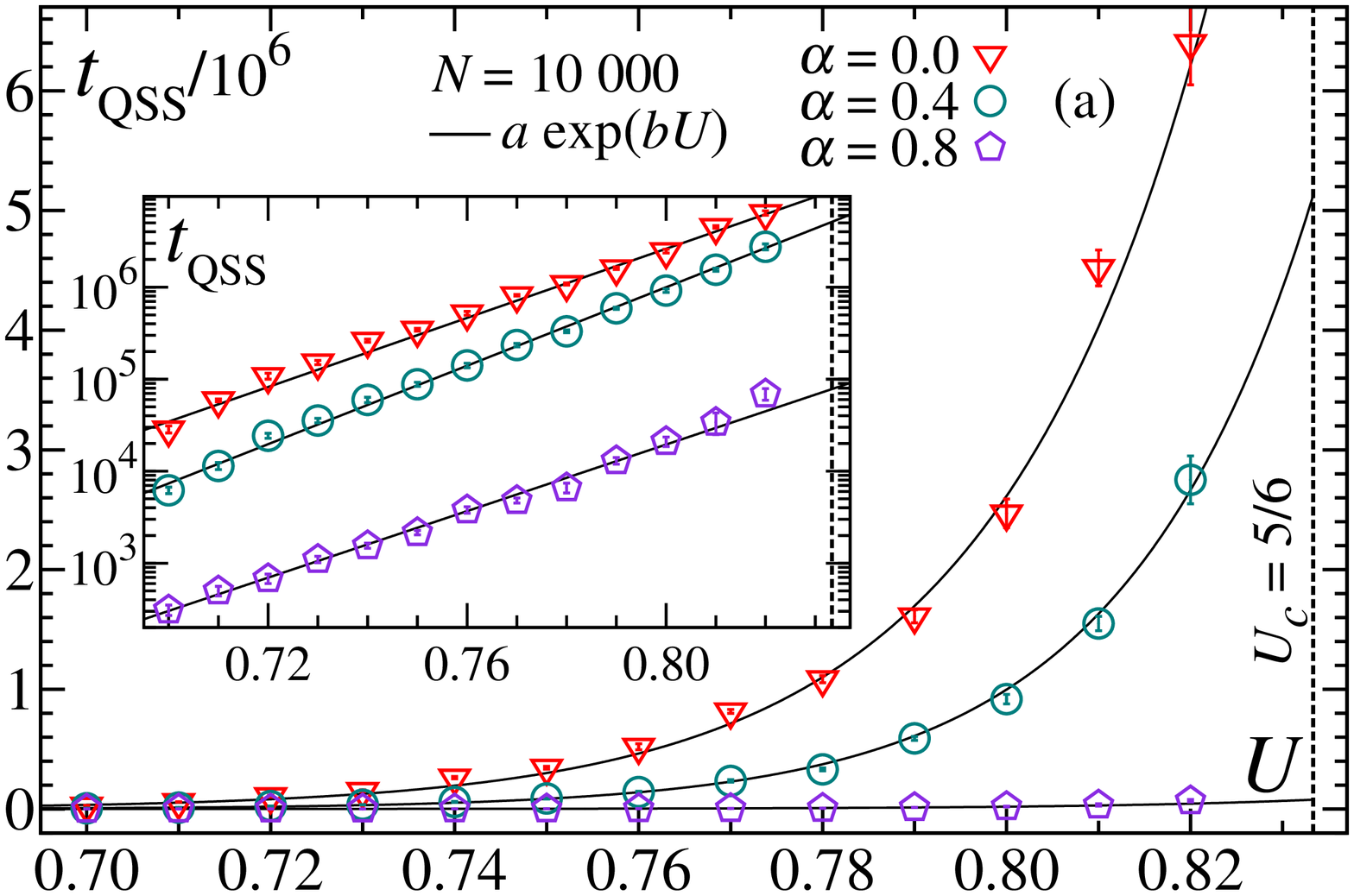}
%\hspace{0.05cm}
\includegraphics[height=0.325\linewidth]{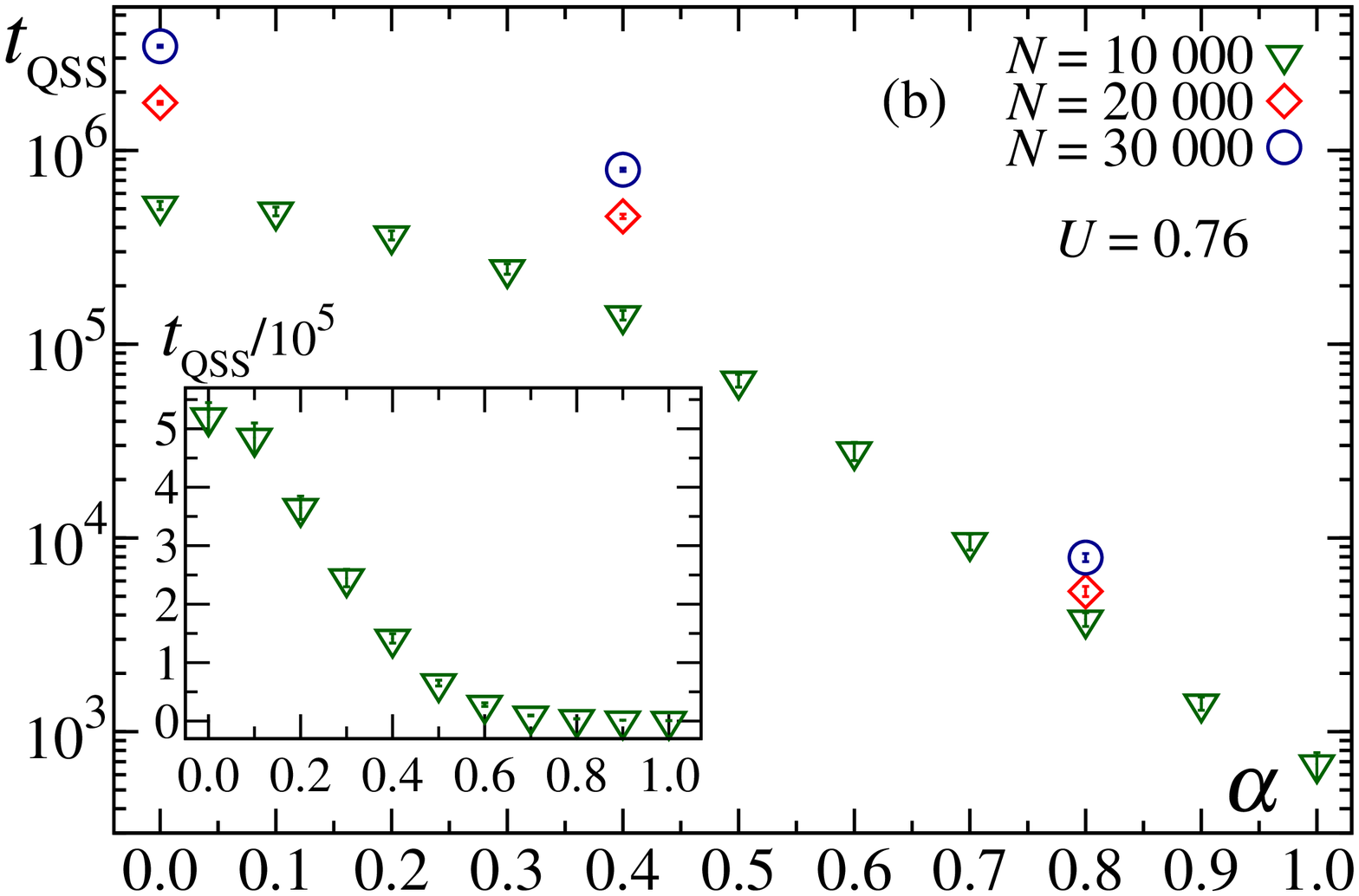}
\caption{\textsf{Behaviours of the lifetime of the QSSs ($t_{\rm QSS}$), with respect 
to the energy~$U$ and to the interaction-range 
parameter~$\alpha$, are shown.  
(a) Fixed number of rotators ($N=10\,000$), three typical values 
of~$\alpha$ ($\alpha=0.0,0.4$, and $0.8$), varying
the energy in the range $0.70 \leq U < U_{c}$~;
the same data are exhibited in a log-linear representation in the inset.  
(b) Fixed energy ($U=0.76$), three values for the total number of 
rotators ($N=10\,000, 20\,000$, and $30\,000$), 
varying~$\alpha$ in the range $0 \leq \alpha \leq 1$; 
the data for $N=10\,000$ are exhibited in a linear-linear representation in the inset.
Symbols correspond to data from numerical simulations of a single copy of 
the Hamiltonian in Eq.~\eqref{eq:Hamiltonian_Heisenberg}, whereas full 
lines  stand for the fits proposed.}}
\label{Fig:Behavior_QSS_Ua}
\end{figure}
%-----------------------------------------------------------------------

The behaviours of the duration~$t_{\rm QSS}$ of the present QSSs with respect to the total energy~$U$, as well as its decrease with the parameter~$\alpha$ are presented in Fig.~\ref{Fig:Behavior_QSS_Ua}.
In figure~\ref{Fig:Behavior_QSS_Ua}(a) we exhibit data of simulations of $N=10\,000$ rotators, for 
three values of~$\alpha$ ($\alpha=0.0,0.4$, and $0.8$), varying the energy in the range $0.70 \leq U < U_{c}$\,. 
These results show that~$t_{\rm QSS}$ grows exponentially for increasing~$U$ in this range, as can be seen more clearly through 
the log-linear representation in the inset. 
Although~$t_{\rm QSS}$ may become very large as $U \to U_{c}$, from the plots of Fig.~\ref{Fig:T_Vs_Tempo_Varios_U}
one sees that the gap between the kinetic temperature of each QSS and the equilibrium state, represented by~$T_{\rm BG}$, 
decreases as~$U$ approaches~$U_{c}$. Hence, in this case, the QSSs disappear 
by means of the difference between these two quantities,
which goes to zero as $U \to U_{c}$, so that 
for $U \gtrsim U_{c}$ one finds no QSSs. 
The dependence of~$t_{\rm QSS}$ on the interaction-range
parameter~$\alpha$ ($0 \leq \alpha \leq 1$) is presented 
in Fig.~\ref{Fig:Behavior_QSS_Ua}(b), where we exhibit data
of simulations for the energy~$U=0.76$ and 
three values for the total number of rotators ($N=10\,000, 20\,000$, and $30\,000$).   
In the log-linear representation, one sees that these
results do strongly depend on~$N$, although they 
are all suggestive that $t_{\rm QSS} \to 0$ as $\alpha \to 1$;
more particularly, the linear-linear representation
of the inset indicates this tendency more clearly in the case $N=10\,000$. 
%-----------------------------------------------------------------------
\section{\label{Sec:Conclusions}Conclusions}
%-----------------------------------------------------------------------
We have carried molecular-dynamics simulations on a one-dimensional Hamiltonian system, composed by~$N$
classical localized Heisenberg rotators on a ring.
We have considered the two-body interaction characterized by 
a distance~$r_{ij}$ between rotators at sites~$i$ and~$j$, decaying with the distance~$r_{ij}$ 
as a power law, $1/r_{ij}^{\alpha}$ ($\alpha \ge 0$).
In this way, one may control the range of the interactions by changing the parameter~$\alpha$, and specially, two well-known
cases may be recovered, namely, the fully-coupled (i.e., mean-field limit) and 
nearest-neighbour-interaction models, in the limits~$\alpha=0$ and $\alpha\to\infty$, respectively. 
For the first time in the literature, we have numerically established 
the validity of the correct scaling for spin dimensionality $n>2$.
Considering this scaling, one obtains the same thermodynamical behaviour for any~$\alpha$ in the
range $0\leq\alpha<1$, as analytically predicted in Ref.~\cite{CampaGiansantiMoroniJPA2003}.

We have investigated the dynamics of the model by following the time evolution of a single copy of 
rotators, for energies~$U$ below its critical value ($U<U_{c}$), considering initial conditions corresponding to zero magnetization. 
Analogously to what happens for the similar model of XY rotators, we have verified the existence 
of quasi-stationary states (QSSs) for values of~$\alpha$ in the range $0 \leq \alpha <1$. 
But, in contrast to the XY rotators, the allowed number of 
initial conditions of the present model is substantially larger.
The particular types of initial conditions that will lead the system 
to robust QSSs was discussed, representing a very relevant matter 
for further analytical investigations.
For a given energy~$U$, our numerical analysis indicated that the 
durations~$t_{\rm QSS}$ grow by increasing~$N$, following a power law, 
$t_{\rm QSS} \sim N^{\gamma}$, where the exponent $\gamma$ gets reduced for 
increasing values of~$\alpha$ in the range $0 \leq \alpha <1$; particularly, our results 
suggest that $\gamma \to 0$ as $\alpha \to 1$. 
The relevance of establishing the growth of $t_{\rm QSS}$ with~$N$ accurately,
achieved herein by considering sufficiently large values of~$N$,
may be assessed if one takes into account the fact that in spite of the 
large amount of effort dedicated to the $\alpha$-XY model, this matter  
is still controversial for this system 
(see, e.g., Refs.~\cite{EttoumiFirpoPRE2013,RochaAmatoFigueiredoPRE2012}).
The particular scaling behaviour, describing the decrease of $\gamma$ 
with respect to $\alpha$, requires further computational efforts 
and is left for future investigations. 
Moreover, for the initial conditions and energy range considered,
we have shown that the duration of the QSSs grows exponentially with the energy, $t_{\rm QSS} \sim \exp(bU)$, for~$N$ and~$\alpha$ fixed. 
Our simulations have indicated that the gaps separating the kinetic temperature of these  QSSs and their corresponding equilibrium
temperatures~$T_{\rm BG}$ become larger for smaller energies, dropping to zero as $U \rightarrow U_{c}$.
In this way, the QSSs disappear by means of the difference between these two quantities,
which goes to zero as~$U \to U_{c}$, so that for $U \gtrsim U_{c}$ one finds no QSSs.

We have found QSSs over a wide range of energies, typically for $0.50 < U < U_{c}$ ($U_{c}=5/6$), although these
states presented zero magnetization throughout a smaller interval, $0.70 \lesssim U < U_{c}$.
It should be mentioned that we have chosen to explore more
deeply the QSSs associated with a given value of energy ($U=0.76$), where such states could 
be observed easily through simulations of a single copy of rotators. However, the results presented herein for this particular
energy should be valid for any similar QSS (characterized by zero magnetization)
in the range $0.70 \lesssim U < U_{c}$.    
By going further below the critical point, roughly for $0.50<U<0.70$, QSSs were still found numerically, but in contrast to the 
above-mentioned ones, they are characterized 
by finite values of magnetization, in spite of the initial conditions of zero magnetization considered.
Within a small energy range (typically $U \lesssim 0.70$), the numerical data of the 
kinetic temperature of these QSSs versus~$U$ yielded a negative microcanonical specific heat.
Although this represents a well-known feature for the HMF model, it is the first time that such a result has been verified for a system of Heisenberg rotators, to our knowledge.

The particular case $\alpha=0.0$ of the present analysis (carried herein for the Cartesian components of angular momenta and spin variables), coincides 
precisely with the one of Ref.~\cite{GuptaMukamelPRE2013}, which have considered angular variables and their corresponding
canonically-conjugated angular momenta.
Herein, we have extended these results for power-law decaying interactions
among spin variables, by investigating situations characterized  by $\alpha>0$, showing that interesting QSSs appear for
any~$\alpha$ in the range $0 \leq \alpha <1$. 
Due to the wide diversity of possibilities for initial conditions
in both spins and angular momenta in the present model, investigations for QSSs under initial conditions different from the ones considered herein would be highly desirable.
%-----------------------------------------------------------------------
% Acknowledgments:
%-----------------------------------------------------------------------
%-----------------------------------------------------------------------
%\section*{Acknowledgments}
%-----------------------------------------------------------------------
\renewcommand{\baselinestretch}{0.99}
\begin{acknowledgments}
\small{We acknowledge useful conversations with C.\ Tsallis, M.\ Jauregui, S.\ T.\ O.\ Almeida, M.\ S.\ Ribeiro, G.\ A.\ Casas, E.\ R.\ P.\ Novais, and  F.\ T.\ L.\ Germani. 
We have benefited from partial financial supports by CNPq, Faperj and Capes (Brazilian funding agencies).}
\end{acknowledgments}
%-----------------------------------------------------------------------
% BIBLIOGRAPHY:
%-----------------------------------------------------------------------
% \section*{References}
% \addcontentsline{toc}{section}{References}
%-----------------------------------------------------------------------

%-----------------------------------------------------------------------
\end{document}